\documentclass[sigconf,nonacm]{acmart}

\setcopyright{none}
\usepackage{placeins}
\usepackage{float}
\begin{document}

\title{Knowledge Distillation across Item and Code Spaces for Parallel Generative Recommendation}
\author{Yixuan Xu}
\email{yxugr@connect.ust.hk}
\affiliation{%
  \institution{The Hong Kong University of Science and Technology}
  \city{Hong Kong}
  \country{China}
}
\author{Hailin Liu}
\email{hliuee@connect.ust.hk}
\affiliation{%
  \institution{The Hong Kong University of Science and Technology}
  \city{Hong Kong}
  \country{China}
}
\author{Linxiang Peng}
\email{plx@mail.ustc.edu.cn}
\affiliation{%
  \institution{University of Science and Technology of China}
  \city{Hefei}
  \country{China}
}
\author{Can Yang}
\email{macyang@ust.hk}
\affiliation{%
  \institution{The Hong Kong University of Science and Technology}
  \city{Hong Kong}
  \country{China}
}
\renewcommand{\shortauthors}{Xu et al.}

\begin{abstract}
Generative recommendation with semantic IDs represents each item as a sequence of discrete codes derived from its semantic representation and predicts these codes to generate recommendations.
Existing semantic-ID recommenders are typically trained with hard labels for the observed next item.
Although such supervision identifies the target, it provides no explicit distributional information about alternative items conditioned on the same user history.
Sequential recommenders can supply this item-level predictive knowledge, but transferring it to semantic-ID generation is challenging because the teacher and student operate in item and code spaces, respectively.
We propose two complementary knowledge-distillation objectives for parallel semantic-ID recommenders, which predict all code positions simultaneously without autoregressive prefix conditioning.
Item Distribution Matching (IDM) reconstructs a catalogue-level item distribution from the student's parallel code predictions and aligns it with the teacher's item distribution.
Code Marginal Matching (CMM) projects the teacher's item probabilities onto each code position using fixed item-to-code assignments, then matches the resulting code marginals to the corresponding student heads.
Both objectives transfer the teacher's soft distributional knowledge while preserving the student's parallel prediction structure.
Experiments on benchmark datasets show that both methods consistently outperform existing state-of-the-art approaches.
On average, IDM improves Recall@10 by 15.39\%, while CMM improves NDCG@10 by 8.28\%.
These results demonstrate that distillation in either reconstructed item space or code-marginal space can substantially strengthen hard-label supervision for parallel generative recommendation.

\end{abstract}



\maketitle

\section{Introduction}
\label{sec:introduction}

Sequential recommendation aims to predict a user's next interaction from their historical behavior. Conventional item-level methods, such as SASRec~\cite{kang2018sasrec}, learn representations of interaction histories and assign scores to individual candidate items~\cite{hidasi2016gru4rec,sun2019bert4rec,zhou2022fmlprec}. Semantic-ID (SID) generative recommenders adopt a different paradigm: each item is represented by a sequence of discrete codes derived from its semantic representation, and recommendation is performed by predicting this code sequence~\cite{wang2024letter,si2024seater,liu2025etegrec,fu2026diger,zheng2024lcrec}. Different items may share the same code at a given position, while their complete SID sequences distinguish them. This structured representation provides a natural target for generative recommendation~\cite{shi2025lladarec,shah2025maskgr}. Early approaches such as TIGER~\cite{rajput2023tiger} predict SID positions autoregressively, conditioning each position on previously generated codes. More recent parallel approaches, including RPG~\cite{hou2025rpg}, predict all positions simultaneously and remove this sequential dependency~\cite{li2026inforid,liu2026flash}. Despite this architectural difference, parallel SID recommenders are still commonly trained with the hard code sequence of a single observed next item.

For a given interaction history, the observed next item induces a one-hot target over the vocabulary of each code position. This target identifies the codes assigned to the observed item, but does not specify how probability mass should be distributed among alternative items or code values. In particular, the fixed item-to-code mapping partitions the catalogue into groups of items that share the same code at each position. A hard target only identifies the group containing the observed item; it provides no graded, history-dependent information about the relative likelihoods of the remaining groups. Consequently, hard-label training does not directly expose the student to the predictive uncertainty encoded in alternative candidates.

A pretrained sequential recommender can provide such complementary supervision through its history-dependent item scores~\cite{tang2018ranking,lee2019collaborative}. Applying a softmax over the candidate catalogue converts these scores into a teacher distribution that assigns different probabilities to alternative items, including items not observed as the next interaction. Existing studies have transferred collaborative or sequential knowledge to generative recommenders through prefix-conditioned token alignment or distillation with learnable identifiers~\cite{lin2026tca4rec,li2026unigrec}. However, these settings differ from a fixed-SID parallel predictor. The teacher produces a distribution over items, whereas the student produces categorical distributions over code values at multiple positions without autoregressive prefix conditioning. The central challenge is therefore how to transfer item-level predictive knowledge into the code distributions of a parallel SID recommender.

The fixed item-to-code mapping provides two natural spaces for this transfer. By combining the student predictions corresponding to an item's complete SID and normalizing the resulting scores over valid catalogue items, the parallel student can induce a catalogue-level item distribution. Alternatively, the teacher's item distribution can be projected onto each code position by summing the probabilities of items that share the same code value. These aggregated probabilities form code-level marginals that can be directly matched to the categorical distributions produced by the corresponding student heads. The two spaces preserve different aspects of the teacher's information: catalogue-level matching retains item-level distinctions, whereas code-level marginal matching provides direct supervision for each parallel prediction head.

Based on these two spaces, we propose two complementary knowledge-distillation objectives, as illustrated in Figure~\ref{fig:item_code_matching}. \emph{Item Distribution Matching} (IDM) combines the student predictions associated with each item's SID, normalizes the resulting scores over the catalogue, and aligns the induced item distribution with the teacher distribution using soft cross-entropy. \emph{Code Marginal Matching} (CMM) instead uses the fixed item-to-code mapping to aggregate teacher item probabilities into a marginal distribution for each code position, and matches these marginals to the corresponding student heads through KL divergence. IDM therefore transfers knowledge through a reconstructed item space, whereas CMM transfers knowledge through a code-marginal space. Both objectives enrich hard-label supervision with history-dependent soft targets while preserving parallel prediction and inference. Our analysis further shows that, under the proposed parameterization, both objectives are driven by teacher-induced code marginals but differ in their student-side normalization: IDM normalizes over catalogue items, whereas CMM normalizes independently within each code head. Empirically, IDM provides stronger improvements in Recall, while CMM yields stronger improvements in NDCG across the evaluated datasets.

Our main contributions are summarized as follows:
\begin{itemize}
    \item We formulate knowledge distillation from item-level teacher distributions to parallel semantic-ID recommenders, addressing the lack of explicit distributional supervision over alternative code values in conventional hard-label training.
    
    \item We propose two complementary objectives for cross-space distillation. IDM matches the teacher distribution in a catalogue-level item space reconstructed from the student's parallel code predictions, whereas CMM projects the teacher distribution into code-position marginals using the fixed item-to-code mapping.
    
    \item We conduct experiments on three benchmark datasets and compare both objectives with state-of-the-art recommendation methods. The proposed methods consistently outperform existing approaches, with IDM achieving an average improvement of 15.39\% in Recall@10 and CMM improving NDCG@10 by 8.28\%.
\end{itemize}
\section{Related Work}
\label{sec:related_work}

\subsection{Sequential Recommendation}
Early sequential recommenders modeled user behavior through personalized Markov transitions, recurrent networks, and convolutional sequence encoders~\cite{rendle2010fpmc,hidasi2016gru4rec,tang2018caser}. With the adoption of self-attention, SASRec and BERT4Rec introduced causal and bidirectional modeling of interaction histories, respectively~\cite{kang2018sasrec,sun2019bert4rec}. Subsequent work has also sought to improve sequence representations through self-supervised and contrastive objectives~\cite{zhou2020s3rec,qiu2022duorec}. In parallel, frequency-domain filtering and hybrid attention have offered alternative ways to capture sequential patterns~\cite{zhou2022fmlprec,du2023fearec}. Despite differences in their architectures and training objectives, these methods learn history-dependent scores for individual candidate items.

\subsection{Generative Recommendation and Semantic-ID Construction}
Beyond encoding interaction histories, generative recommendation also revisits how recommendation tasks are formulated. For instance, P5 casts multiple recommendation tasks as text generation~\cite{geng2022p5}, whereas HSTU models user actions through sequential transduction~\cite{zhai2024hstu}. Further developments include hierarchical item--user language modeling, session-wise generation that unifies retrieval and ranking, and the integration of cross features into sequential transducers~\cite{chen2024hllm,deng2025onerec,han2025mtgr}. Together, these approaches illustrate the range of architectures and prediction targets explored in generative recommendation.

Alongside these developments, another line of research has focused on representing catalogue items through discrete codes drawn from shared codebooks. Vector and product quantization have been used to construct item representations for sequential recommendation~\cite{hou2023vqrec,petrov2024recjpq}. In generative retrieval, TIGER instead uses content-derived semantic IDs (SIDs) as autoregressive prediction targets~\cite{rajput2023tiger}. Subsequent work has refined identifier construction by incorporating language and collaborative signals, semantic tree structures, and diverse code assignments~\cite{zheng2024lcrec,si2024seater,wang2024letter}. More recently, ETEGRec and DIGER have connected identifier learning more closely to recommendation optimization~\cite{liu2025etegrec,fu2026diger}, while ActionPiece explores context-dependent tokenization of action sequences~\cite{hou2025actionpiece}. Given such identifiers, a separate modeling choice concerns how their code positions are predicted.

\subsection{Parallel Semantic-ID Recommendation}
Autoregressive SID recommenders generate code positions sequentially, conditioning each prediction on preceding codes. To remove this dependency, RPG predicts long, unordered identifiers through parallel heads and uses graph-guided decoding to recover valid catalogue items~\cite{hou2025rpg}. AsymRec also adopts parallel code prediction with continuous item inputs~\cite{huang2026asymrec}. In both architectures, the prediction heads share a history representation without conditioning on previously generated code values, allowing all positions to be evaluated together. Within this paradigm, recent studies have examined the identifiers themselves, including semantic capacity allocation across positions and SimHash-based construction~\cite{li2026inforid,liu2026flash}. An alternative non-autoregressive approach progressively refines predicted codes: LLaDA-Rec uses discrete diffusion with an adaptive decoding order, whereas MaskGR iteratively reconstructs masked positions~\cite{shi2025lladarec,shah2025maskgr}. Unlike these iterative schemes, single-pass predictors produce position-wise categorical distributions in one model evaluation. These distributions are combined to score catalogue items, while standard training uses the observed next item's assigned codes as hard targets without specifying graded probabilities for alternative items.

\subsection{Knowledge Distillation for Recommendation}
Alongside advances in identifier construction and decoding, knowledge distillation transfers predictive information from a teacher to a student~\cite{hinton2015distilling}. In conventional recommendation, this has included transferring ranking preferences and collaborative knowledge~\cite{tang2018ranking,lee2019collaborative}, as well as relations among learned representations~\cite{kang2021topology}. For generative recommenders, knowledge transfer has also been adapted to the student's identifier structure and token predictions. UniGRec transfers collaborative information into its tokenizer and recommender~\cite{li2026unigrec}, while SmartGR incorporates SID hierarchy and beam-level ranking~\cite{zhang2026smartgr}. TCA4Rec introduces item-level collaborative supervision into autoregressive token prediction~\cite{lin2026tca4rec}. We formulate item-space and code-marginal distillation for parallel SID predictors and analyze how catalogue-level and head-wise normalization lead to different optimization signals.

\section{Methodology}
\label{sec:methodology}

We present two distribution-matching objectives that transfer item-level predictive knowledge to a parallel semantic-ID (SID) recommender. As illustrated in Figure~\ref{fig:item_code_matching}, the fixed item-to-code mapping supports both catalogue-level matching in IDM and position-wise code matching in CMM. We first establish the prediction and ranking interface, then introduce each objective and analyze the resulting optimization differences before describing training.

\subsection{Problem Formulation}
\label{sec:method_formulation}

\paragraph{Parallel SID prediction and ranking.}
Let $\mathcal I=\{1,\ldots,N\}$ denote the valid item catalogue and $(x,y)$ a training instance comprising an interaction history $x$ and its observed next item $y$. Each item $i$ has a fixed SID $\mathbf z_i=(z_{i,1},\ldots,z_{i,H})$, where $z_{i,h}\in\mathcal C=\{0,\ldots,K-1\}$. A parallel recommender parameterized by $\theta$ encodes $x$ into a shared representation and simultaneously predicts all $H$ code positions without autoregressive prefix conditioning. For head $h$, let $\mathbf u_h(x)\in\mathbb R^K$ denote its unscaled outputs and $\boldsymbol\ell_h(x)=\mathbf u_h(x)/\tau_S$ its effective logits at student temperature $\tau_S>0$. The resulting categorical distribution is
\begin{equation}
q_h(c\mid x)
=\frac{\exp(\ell_{h,c}(x))}
{\sum_{k\in\mathcal C}\exp(\ell_{h,k}(x))}.
\label{eq:head_distribution}
\end{equation}
To rank candidate items from these parallel predictions, the student aggregates the log-probabilities of each item's assigned codes:
\begin{equation}
\widetilde s_i(x)=\frac1H\sum_{h=1}^{H}\log q_h(z_{i,h}\mid x),
\label{eq:inference_score}
\end{equation}
where $\widetilde s_i(x)$ is the ranking score for candidate $i$. Because the head-wise log-partition terms in Eq.~\eqref{eq:head_distribution} contribute an identical history-dependent constant across candidates, this score is equivalent to averaging the assigned logits for both ranking and catalogue softmax normalization. This equivalence provides the scoring interface for IDM (Section~\ref{sec:idm}; proof in Appendix~\ref{app:equivalent_scores}).

\paragraph{Hard labels and distributional supervision.}
Standard training uses the codes of the observed next item as hard targets for all prediction heads:
\begin{equation}
\mathcal L_{\mathrm{hard}}(x,y)
=-\frac1H\sum_{h=1}^{H}\log q_h(z_{y,h}\mid x).
\label{eq:hard_loss}
\end{equation}
Each observation identifies one code per position but leaves the predictive probabilities of alternative catalogue items unspecified. To supply this graded supervision, we use a pretrained SASRec~\cite{kang2018sasrec} as a frozen item-level teacher. Given its scores $a_i(x)$, we form a history-dependent distribution over the catalogue:
\begin{equation}
p_T(i\mid x)
=\frac{\exp(a_i(x)/\tau_T)}
{\sum_{j\in\mathcal I}\exp(a_j(x)/\tau_T)},
\label{eq:teacher_distribution}
\end{equation}
where $\tau_T>0$ controls the concentration of the teacher probabilities. The teacher thus predicts probabilities over items, whereas the student directly predicts categorical distributions over code values.

\begin{figure*}[t]
  \centering
  \includegraphics[width=\textwidth]{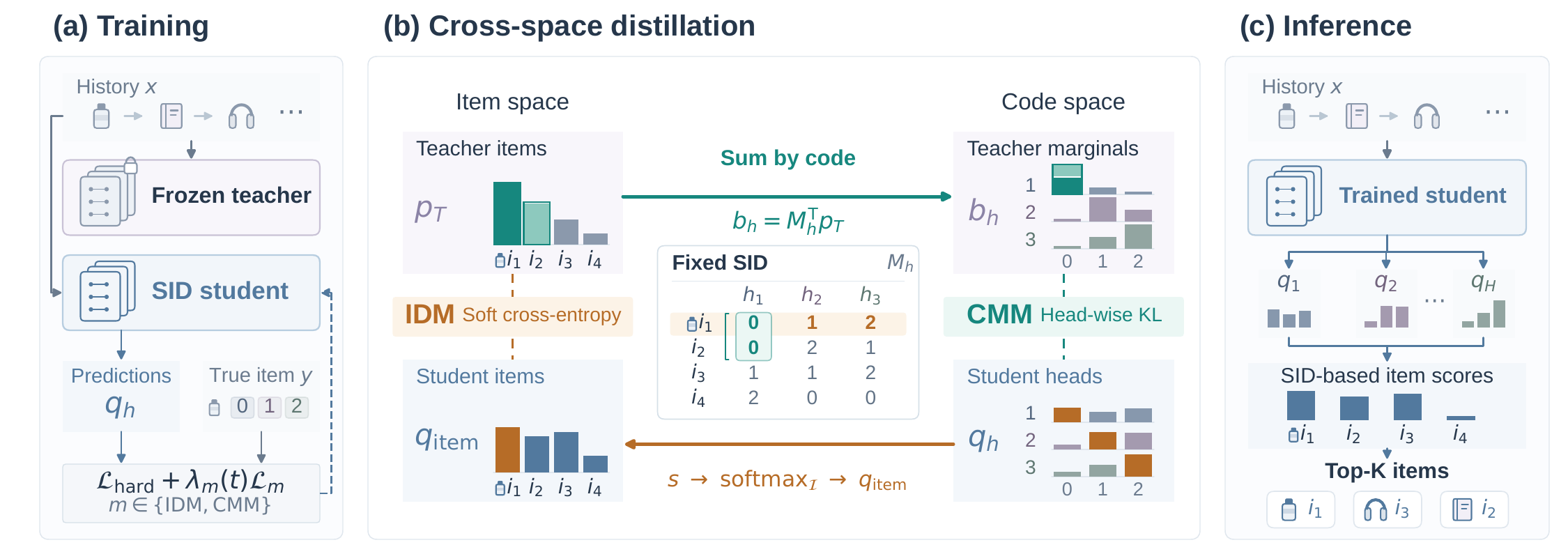}
  \caption{Overview of item- and code-space knowledge distillation for parallel semantic-ID recommendation.
  (a) A parallel SID student learns from hard labels and a frozen teacher using either IDM or CMM.
  (b) Fixed assignments $M_h$ connect teacher item probabilities with student code predictions. IDM matches distributions over valid catalogue items; CMM matches teacher-derived code marginals with student head distributions. 
  (c) At inference, the student predicts SID codes in parallel and ranks valid items without the teacher.}
  \Description{Three panels show the training, cross-space distillation, and inference. A frozen teacher produces item probabilities and a trainable parallel SID student produces code logits. Fixed item-to-code assignments gather student logits into catalogue scores for IDM and aggregate teacher probabilities into code marginals for CMM. Each objective is trained separately with hard-label supervision. At inference, the student produces catalogue scores and selects the top-ranked items.}
  \label{fig:item_code_matching}
\end{figure*}

\paragraph{Cross-space mapping interface.}
We connect the two prediction spaces through deterministic SID assignments. For each position $h$, define the assignment matrix $M_h\in\{0,1\}^{N\times K}$ by $(M_h)_{i,c}=\mathbf 1[z_{i,h}=c]$, so row $i$ indexes the code assigned to item $i$. With the teacher probabilities collected in $\mathbf p_T(x)\in\mathbb R^N$, these matrices connect the spaces in both directions: $M_h$ gathers student code logits into catalogue scores for IDM (Section~\ref{sec:idm}), while $M_h^\top$ projects teacher item probabilities onto position-wise marginals for CMM (Section~\ref{sec:cmm}), as illustrated in Figure~\ref{fig:item_code_matching}.

\subsection{Item Distribution Matching}
\label{sec:idm}

Item Distribution Matching (IDM) uses the student's catalogue scoring function to match the teacher's preferences in item space. The mean-logit scores equivalent to Eq.~\eqref{eq:inference_score} are
\begin{equation}
\begin{aligned}
s_i(x)&=\frac1H\sum_{h=1}^{H}\ell_{h,z_{i,h}}(x),\\
\mathbf s(x)&=\frac1H\sum_{h=1}^{H}M_h\boldsymbol\ell_h(x).
\end{aligned}
\label{eq:item_code_mapping}
\end{equation}
Here, $M_h$ gathers the logit assigned to each item at position $h$, and averaging across heads combines their contributions into a single item score.

Applying a softmax over the catalogue to the scores in Eq.~\eqref{eq:item_code_mapping} gives
\begin{equation}
\begin{aligned}
q_{\mathrm{item}}(i\mid x)
&=\frac{\exp(s_i(x))}
{\sum_{j\in\mathcal I}\exp(s_j(x))}\\
&=\frac{\prod_{h=1}^{H}q_h(z_{i,h}\mid x)^{1/H}}
{\sum_{j\in\mathcal I}\prod_{h=1}^{H}q_h(z_{j,h}\mid x)^{1/H}}.
\end{aligned}
\label{eq:idm_catalogue_distribution}
\end{equation}
The second expression makes the connection to parallel code prediction explicit: each item's unnormalized probability is proportional to the geometric mean of its assigned code probabilities. Catalogue normalization then places all valid items in a common probability space, coupling the contributions of different heads through their combined item scores.

We align the teacher and student distributions by minimizing the soft cross-entropy
\begin{equation}
\mathcal L_{\mathrm{IDM}}(x)
=-\sum_{i\in\mathcal I}p_T(i\mid x)
\log q_{\mathrm{item}}(i\mid x).
\label{eq:idm_loss}
\end{equation}
IDM supervises the combined item scores used for ranking. Since multiple items can share a code, changing \(\ell_{h,c}\) shifts the scores of all items assigned to code \(c\) at position \(h\). The catalogue softmax couples these score changes through a shared normalizer, allowing the item-level objective to jointly train the parallel prediction heads.

\subsection{Code Marginal Matching}
\label{sec:cmm}

Whereas IDM supervises the combined item predictions through normalization over $N$ catalogue items, a complementary approach is to supervise each parallel head directly in its $K$-way categorical space. We therefore propose Code Marginal Matching (CMM), which uses the item partitions induced by each code position to project teacher knowledge into head-specific marginal targets.

At position $h$, let $\mathcal I_{h,c}=\{i\in\mathcal I:z_{i,h}=c\}$ contain the items assigned to code $c$. Because each item has exactly one code at that position, the nonempty groups $\mathcal I_{h,c}$ partition the catalogue. The transpose $M_h^\top$ aggregates teacher probabilities within these groups, giving
\begin{equation}
\begin{aligned}
\mathbf b_h(x)&=M_h^\top\mathbf p_T(x),\\
b_h(c\mid x)
&=\sum_{i\in\mathcal I_{h,c}}p_T(i\mid x)\\
&=\Pr_{J\sim p_T(\cdot\mid x)}[z_{J,h}=c].
\end{aligned}
\label{eq:cmm_marginal}
\end{equation}
This projection defines the exact marginal of the teacher's item distribution under the fixed SID mapping. Its probabilities sum to one, since each item contributes to one code at position $h$, while codes unused by the catalogue receive zero teacher mass. Thus, the target $b_h(\cdot\mid x)$ and student prediction $q_h(\cdot\mid x)$ are distributions over the same code vocabulary.

The position-wise partitions also explain why a single teacher distribution can supervise every parallel head. Two items may share a code at one position but have different codes at another; their probability mass is aggregated in the former head and separated in the latter. Consequently, the teacher's history-dependent item probabilities can induce different marginal targets across SID positions. All of these targets can be constructed simultaneously, matching the student's parallel prediction structure.

CMM aligns each head-wise categorical distribution with its corresponding teacher marginal through KL divergence:
\begin{equation}
\mathcal L_{\mathrm{CMM}}(x)
=\frac1H\sum_{h=1}^{H}
D_{\mathrm{KL}}\!\left(
 b_h(\cdot\mid x)\,\Vert\,q_h(\cdot\mid x)
\right).
\label{eq:cmm_loss}
\end{equation}
The loss directly supervises each head with its teacher marginal, and gradients from all heads update the shared encoder. As in \(\mathcal L_{\mathrm{hard}}\), averaging over positions gives each head equal weight.

\subsection{Relationship between Matching Objectives}
\label{sec:matching_relationship}

IDM and CMM transfer the same teacher knowledge through different student normalizations. We characterize CMM in factorized code space, then relate the objectives and their gradients. Proofs are given in Appendix~\ref{app:mathematical_details}.

\paragraph{Factorized code-distribution matching.}
To understand CMM from a joint-distribution perspective, consider the distributions over complete code tuples $\mathbf z\in\mathcal C^H$ induced by the teacher and student. The teacher aggregates probabilities for identical SIDs, while the parallel student defines a product distribution:
\begin{equation}
\begin{aligned}
P_T^Z(\mathbf z\mid x)
&=\sum_{i:\mathbf z_i=\mathbf z}p_T(i\mid x),\\
Q_\theta^Z(\mathbf z\mid x)
&=\prod_{h=1}^{H}q_h(z_h\mid x).
\end{aligned}
\label{eq:joint_code_distributions}
\end{equation}
The following proposition separates teacher dependence across code positions from the terms optimized by the student.

\begin{proposition}[Factorized forward-KL decomposition]
\label{prop:factorized_forward_kl}
For a fixed teacher distribution and SID mapping, the forward KL from $P_T^Z$ to the factorized student distribution $Q_\theta^Z$ satisfies
\begin{equation}
D_{\mathrm{KL}}(P_T^Z\Vert Q_\theta^Z)
=\operatorname{TC}(P_T^Z)+H\mathcal L_{\mathrm{CMM}},
\label{eq:factorized_forward_kl}
\end{equation}
where $\operatorname{TC}(P_T^Z)=D_{\mathrm{KL}}(P_T^Z\Vert\prod_h b_h)$ is the total correlation of the teacher-induced code distribution.
\end{proposition}
Since the total-correlation term is independent of student parameters, CMM minimizes exactly the student-dependent part of the joint forward KL. In the unrestricted product family, the optimum is $q_h=b_h$ for every head: matching the teacher's joint code distribution therefore reduces to aligning its position-wise marginals with the parallel predictor.

\paragraph{Normalization and gradients.}
While Proposition~\ref{prop:factorized_forward_kl} characterizes CMM over complete code tuples, IDM normalizes over valid catalogue items. Expanding both losses reveals their shared linear scoring term, $-\frac1H\sum_{h,c}b_h(c\mid x)\ell_{h,c}(x)$. Apart from a student-independent teacher-entropy term, the losses differ through their catalogue-level and head-wise log-partition normalizers:
\begin{equation}
\Delta_{\mathrm{norm}}
=\log\sum_{i\in\mathcal I}\exp(s_i)
-\frac1H\sum_{h=1}^{H}\log\sum_{c\in\mathcal C}\exp(\ell_{h,c}),
\label{eq:partition_difference}
\end{equation}
where conditioning on $x$ is omitted. This yields the following exact relationship.

\begin{proposition}[Cross-space loss and gradient identity]
\label{prop:cross_space_identity}
Assume the teacher distribution and SID mapping are fixed. With the mean-logit item scores in Eq.~\eqref{eq:item_code_mapping} and unit catalogue temperature, IDM and CMM satisfy
\begin{equation}
\begin{aligned}
\mathcal L_{\mathrm{IDM}}
&=\mathcal L_{\mathrm{CMM}}
+\frac1H\sum_{h=1}^{H}\mathcal H(b_h)
+\Delta_{\mathrm{norm}},\\
\nabla_\theta\mathcal L_{\mathrm{IDM}}
&=\nabla_\theta\mathcal L_{\mathrm{CMM}}
+\nabla_\theta\Delta_{\mathrm{norm}},
\end{aligned}
\label{eq:normalization_identity}
\end{equation}
where $\mathcal H(b_h)$ is the entropy of the teacher marginal at head $h$.
\end{proposition}

The core optimization insight of Proposition~\ref{prop:cross_space_identity} is that identical teacher marginals can produce different updates through student-side normalization. The teacher entropy contributes no student gradient; defining the catalogue-induced marginal $\bar{\mathbf q}_h=M_h^\top\mathbf q_{\mathrm{item}}$ makes this distinction explicit at the logit level:
\begin{equation}
\begin{aligned}
\frac{\partial\mathcal L_{\mathrm{IDM}}}{\partial\ell_{h,c}}
&=\frac{\bar q_h(c)-b_h(c)}{H},\\
\frac{\partial\mathcal L_{\mathrm{CMM}}}{\partial\ell_{h,c}}
&=\frac{q_h(c)-b_h(c)}{H}.
\end{aligned}
\label{eq:matching_logit_gradients}
\end{equation}
Both gradients contain the teacher term $-b_h(c)/H$, but compare it with different student probabilities: CMM uses the per-head distribution $q_h(c)$, whereas IDM uses $\bar q_h(c)$, which depends on item scores combined across all heads. Thus, CMM adjusts each head toward its teacher marginal directly, while IDM adjusts the marginal induced by the catalogue ranking distribution. The normalization space determines how the same teacher knowledge guides the parallel predictor.

\subsection{Training Objective and Gradient Calibration}
\label{sec:method_optimization}

We train IDM and CMM separately, combining each auxiliary loss with the original hard-label supervision. For a minibatch $\mathcal B$, let $t$ count the optimizer updates completed before the current batch. The training objective is
\begin{equation}
\mathcal J_m(\theta;t)
=\frac1{|\mathcal B|}\sum_{(x,y)\in\mathcal B}
\!\left[
\mathcal L_{\mathrm{hard}}(x,y)
+\lambda_m w(t)\mathcal L_m(x)
\right],
\label{eq:training_objective}
\end{equation}
where $m\in\{\mathrm{IDM},\mathrm{CMM}\}$ and $w(t)$ gradually introduces the auxiliary supervision. Hard labels and soft distributions allocate prediction error differently across the $N$ catalogue items or $K$ code classes, yielding objective-dependent gradient scales. We account for these scales through $\lambda_m=r_mC_m$: $r_m$ controls nominal distillation strength, and $C_m$ is the median ratio of hard-label to auxiliary $L_2$ gradient norms across early calibration batches. Measuring the norms over all trainable parameters balances the two signals at the level of the optimized student. Appendix~\ref{app:calibration} details the calibration and ramp.

\section{Experiments}
\label{sec:experiments}
We organize the experiments around four research questions:
\begin{itemize}
    \item \textbf{RQ1 (Overall performance):} How do IDM and CMM compare with hard-label training and existing recommenders?
    \item \textbf{RQ2 (Ablation):} How do soft targets and code-head coverage contribute to performance?
    \item \textbf{RQ3 (Sensitivity):} How do teacher data, temperature, and distillation strength affect knowledge transfer?
    \item \textbf{RQ4 (Combination):} How does joint training compare with score fusion?
\end{itemize}

\subsection{Experimental Setup}
\label{sec:exp-setup}
\paragraph{Datasets and splits}
We use the Beauty, Toys and Games (Toys), and Sports and Outdoors (Sports) categories of Amazon Reviews 2014. Table~\ref{tab:exp-data} summarizes the 5-core datasets. For each user, the last interaction is held out for Test and the penultimate interaction for Validation; earlier interactions provide training supervision. Histories are truncated to at most 50 items. Appendix~\ref{app:implementation} details the training-window construction and supervised positions.

\begin{table}[!htbp]
\centering
\caption{Dataset statistics. Each user contributes one Validation target and one Test target.}
\label{tab:exp-data}
\small
\setlength{\tabcolsep}{2pt}
\begin{tabular*}{\linewidth}{@{\extracolsep{\fill}}lrrr@{}}
\toprule
Statistic & Beauty & Toys & Sports \\
\midrule
Users & 22,363 & 19,412 & 35,598 \\
Items & 12,101 & 11,924 & 18,357 \\
Interactions & 198,502 & 167,597 & 296,337 \\
Training targets & 131,413 & 109,361 & 189,543 \\
\bottomrule
\end{tabular*}
\end{table}

\paragraph{Models and comparisons}
We compare against two families of sequential recommenders. \emph{Item-ID-based methods} include Caser~\cite{tang2018caser}, GRU4Rec~\cite{hidasi2016gru4rec}, HGN~\cite{ma2019hgn}, BERT4Rec~\cite{sun2019bert4rec}, SASRec~\cite{kang2018sasrec}, FDSA~\cite{zhang2019fdsa}, and $\mathrm{S}^{3}$-Rec~\cite{zhou2020s3rec}. \emph{Generative and SID-based methods} comprise the quantization-based RecJPQ~\cite{petrov2024recjpq} and VQ-Rec~\cite{hou2023vqrec}, autoregressive TIGER~\cite{rajput2023tiger} and HSTU~\cite{zhai2024hstu}, and parallel SID predictors RPG~\cite{hou2025rpg} and AsymRec~\cite{huang2026asymrec}.

Our student follows AsymRec, with 16 quantization subspaces and three residual levels per subspace, yielding $16\times3=48$ heads with 256 codes each. Hard-label AsymRec provides the controlled baseline, and pretrained SASRec supplies teacher probabilities. We reproduce RPG using its released configuration and graph-guided decoder; Table~\ref{tab:exp-main} identifies the sources of all comparison results.

\begin{table*}[!t]
\centering
\caption{Overall Test results. R and N denote Recall and NDCG. Results for RPG, AsymRec, IDM, and CMM are averaged over three runs; all other baseline results are taken from RPG~\cite{hou2025rpg}, Table 2. Bold and underline mark the best and second-best values in each metric column. Gain is the relative improvement of the best IDM/CMM mean over hard-label AsymRec. Significance testing: $*$ indicates statistically significant improvements of IDM/CMM over AsymRec via a two-sided paired $t$-test over users ($p<0.05$), while $-$ denotes no significant difference.}
\label{tab:exp-main}
\label{tab:rpg-published-full}
\label{tab:app-rpg-paper}
\footnotesize
\setlength{\tabcolsep}{2pt}
\begin{tabular*}{\linewidth}{@{\extracolsep{\fill}}lrrrrrrrrrrrr@{}}
\toprule
Method & \multicolumn{4}{c}{Beauty} & \multicolumn{4}{c}{Toys} & \multicolumn{4}{c}{Sports} \\
\cmidrule(lr){2-5}\cmidrule(lr){6-9}\cmidrule(lr){10-13}
 & R@5 & R@10 & N@5 & N@10 & R@5 & R@10 & N@5 & N@10 & R@5 & R@10 & N@5 & N@10 \\
\midrule
\multicolumn{13}{l}{\emph{Item-ID-based}} \\
Caser & 0.0205 & 0.0347 & 0.0131 & 0.0176 & 0.0166 & 0.0270 & 0.0107 & 0.0141 & 0.0116 & 0.0194 & 0.0072 & 0.0097 \\
GRU4Rec & 0.0164 & 0.0283 & 0.0099 & 0.0137 & 0.0097 & 0.0176 & 0.0059 & 0.0084 & 0.0129 & 0.0204 & 0.0086 & 0.0110 \\
HGN & 0.0325 & 0.0512 & 0.0206 & 0.0266 & 0.0321 & 0.0497 & 0.0221 & 0.0277 & 0.0189 & 0.0313 & 0.0120 & 0.0159 \\
BERT4Rec & 0.0203 & 0.0347 & 0.0124 & 0.0170 & 0.0116 & 0.0203 & 0.0071 & 0.0099 & 0.0115 & 0.0191 & 0.0075 & 0.0099 \\
SASRec & 0.0387 & 0.0605 & 0.0249 & 0.0318 & 0.0463 & 0.0675 & 0.0306 & 0.0374 & 0.0233 & 0.0350 & 0.0154 & 0.0192 \\
FDSA & 0.0267 & 0.0407 & 0.0163 & 0.0208 & 0.0228 & 0.0381 & 0.0140 & 0.0189 & 0.0182 & 0.0288 & 0.0122 & 0.0156 \\
$\mathrm{S}^{3}$-Rec & 0.0387 & 0.0647 & 0.0244 & 0.0327 & 0.0443 & 0.0700 & 0.0294 & 0.0376 & 0.0251 & 0.0385 & 0.0161 & 0.0204 \\
\midrule
\multicolumn{13}{l}{\emph{Generative and SID-based}} \\
RecJPQ & 0.0311 & 0.0482 & 0.0167 & 0.0222 & 0.0331 & 0.0484 & 0.0182 & 0.0231 & 0.0141 & 0.0220 & 0.0076 & 0.0102 \\
VQ-Rec & 0.0457 & 0.0664 & 0.0317 & 0.0383 & 0.0497 & 0.0737 & 0.0346 & 0.0423 & 0.0208 & 0.0300 & 0.0144 & 0.0173 \\
TIGER & 0.0454 & 0.0648 & 0.0321 & 0.0384 & 0.0521 & 0.0712 & 0.0371 & 0.0432 & 0.0264 & 0.0400 & 0.0181 & 0.0225 \\
HSTU & 0.0469 & 0.0704 & 0.0314 & 0.0389 & 0.0433 & 0.0669 & 0.0281 & 0.0357 & 0.0258 & 0.0414 & 0.0165 & 0.0215 \\
RPG & $0.0537$ & $0.0798$ & $0.0370$ & $0.0454$ & $0.0579$ & $0.0885$ & $0.0387$ & $0.0485$ & $0.0292$ & $0.0435$ & $0.0203$ & $0.0249$ \\
AsymRec & $0.0645$ & $0.0940$ & $0.0452$ & $0.0547$ & $0.0702$ & $0.1002$ & $0.0489$ & $0.0586$ & $0.0382$ & $0.0573$ & $0.0260$ & $0.0322$ \\
IDM (ours) & $\mathbf{0.0719}^{*}$ & $\mathbf{0.1094}^{*}$ & $\underline{0.0468}^{*}$ & $\underline{0.0590}^{*}$ & $\mathbf{0.0804}^{*}$ & $\mathbf{0.1176}^{*}$ & $\underline{0.0518}^{*}$ & $\underline{0.0638}^{*}$ & $\mathbf{0.0423}^{*}$ & $\mathbf{0.0644}^{*}$ & $\underline{0.0262}^{-}$ & $\underline{0.0333}^{*}$ \\
CMM (ours) & $\underline{0.0714}^{*}$ & $\underline{0.1069}^{*}$ & $\mathbf{0.0478}^{*}$ & $\mathbf{0.0592}^{*}$ & $\underline{0.0784}^{*}$ & $\underline{0.1144}^{*}$ & $\mathbf{0.0532}^{*}$ & $\mathbf{0.0647}^{*}$ & $\underline{0.0422}^{*}$ & $\underline{0.0639}^{*}$ & $\mathbf{0.0272}^{*}$ & $\mathbf{0.0341}^{*}$ \\
\midrule
Gain vs. AsymRec & +11.41\% & +16.42\% & +5.69\% & +8.33\% & +14.60\% & +17.36\% & +8.61\% & +10.44\% & +10.87\% & +12.39\% & +4.42\% & +6.08\% \\
\bottomrule
\end{tabular*}
\end{table*}

\paragraph{Evaluation}
We report Recall and NDCG at cutoffs 5 and 10 (R@5, R@10, N@5, and N@10). AsymRec, IDM, and CMM rank the full item catalogue without sampled negatives or historical-item masking. RPG retrieves from the same item catalogue using its released graph decoder. Results for our reproduced baselines (RPG and AsymRec) and proposed methods are averaged over three random seeds.

\paragraph{Training and selection}
Students use AdamW with batch size 256 for up to 150 epochs, with 10,000 learning-rate warm-up updates followed by cosine decay and gradient clipping at norm 1. We select checkpoints by Validation N@10 with patience 20 and choose learning rates and distillation strengths on Validation. To ensure fair comparisons, all student variants share identical SID assignments, pretrained teacher checkpoints, and random seed initializations. Each selected hyperparameter configuration and calibrated coefficient is reused across all three runs. Appendix~\ref{app:calibration} provides the calibration procedure, ramp schedule, search grids, and selected values.

\subsection{Overall Performance (RQ1)}
\label{sec:exp-main}

\paragraph{Comparison with existing methods}
Table~\ref{tab:exp-main} shows that IDM and CMM achieve higher mean scores than all compared baselines across the three datasets and four metrics. Both improve upon AsymRec, the strongest baseline in the table, with gains at both cutoffs 5 and 10. These results demonstrate the effectiveness of distribution matching across datasets and retrieval depths.

\paragraph{Effect of distributional supervision}
Relative to hard-label AsymRec, IDM improves R@10 by 15.39\% on average, while CMM improves N@10 by 8.28\%. This comparison retains the student architecture, SID assignments, and inference rule, demonstrating the effectiveness of enriching supervision within the same parallel predictor. The teacher's history-dependent probabilities complement the observed next-item labels by assigning graded targets to alternative candidates and their associated codes.

\paragraph{IDM versus CMM}
IDM achieves higher Recall and CMM higher NDCG at both cutoffs across all three datasets, revealing complementary retrieval and ordering behavior. The gradients in Eq.~\eqref{eq:matching_logit_gradients} explain how the normalization space changes the learning signal. IDM allocates probability only to valid catalogue items and couples all heads through competition among their combined item scores, directing supervision toward recovering plausible candidates in the top-$K$ set. CMM instead aligns each head directly with the teacher's code marginal, reinforcing shared code patterns supported by high-probability items without reconstructing their joint catalogue scores. The observed Recall--NDCG split is consistent with this distinction between global candidate competition and direct code-level alignment: the former favors candidate coverage, while the latter favors precise ordering near the top of the ranked list.

\FloatBarrier

\subsection{Ablation Study (RQ2)}
\label{sec:exp-ablation}

To examine the source of these gains, Table~\ref{tab:exp-ablation} separates two aspects of supervision: the probability information in teacher targets and its coverage across code heads. To compare their supervision signals, we calibrate each variant separately. Appendix~\ref{app:ablations} gives their definitions, calibration settings, and additional metrics.

\begin{table}[!htbp]
\centering
\caption{Ablation results (mean performance over three random seeds). Variants change only the auxiliary targets or supervised code heads; hard-label supervision covers all 48 heads. Bold denotes the highest mean within each objective group and metric.}
\label{tab:exp-ablation}
\small
\setlength{\tabcolsep}{2pt}
\begin{tabular*}{\linewidth}{@{\extracolsep{\fill}}lrrrrrr@{}}
\toprule
Variant & \multicolumn{2}{c}{Beauty} & \multicolumn{2}{c}{Toys} & \multicolumn{2}{c}{Sports} \\
 & R@10 & N@10 & R@10 & N@10 & R@10 & N@10 \\
\midrule
IDM (full) & $\mathbf{0.1094}$ & $\mathbf{0.0590}$ & $\mathbf{0.1176}$ & $\mathbf{0.0638}$ & $\mathbf{0.0644}$ & $\mathbf{0.0333}$ \\
Top-1 item & $0.1041$ & $0.0548$ & $0.1141$ & $0.0608$ & $0.0606$ & $0.0294$ \\
Uniform-smoothed item & $0.1038$ & $0.0548$ & $0.1144$ & $0.0612$ & $0.0602$ & $0.0293$ \\
\midrule
CMM (full) & $\mathbf{0.1069}$ & $\mathbf{0.0592}$ & $\mathbf{0.1144}$ & $\mathbf{0.0647}$ & $\mathbf{0.0639}$ & $0.0341$ \\
Hard code & $0.1010$ & $0.0545$ & $0.1119$ & $0.0616$ & $0.0595$ & $0.0292$ \\
Uniform-smoothed code & $0.1006$ & $0.0547$ & $0.1103$ & $0.0617$ & $0.0599$ & $0.0305$ \\
16 heads (level 1) & $0.1011$ & $0.0584$ & $0.1095$ & $0.0639$ & $0.0613$ & $0.0346$ \\
32 heads (levels 1--2) & $0.1046$ & $0.0591$ & $0.1112$ & $0.0643$ & $0.0635$ & $\mathbf{0.0347}$ \\
\bottomrule
\end{tabular*}
\end{table}

\paragraph{Information in soft targets}
The hard-target variants retain only the teacher's top-ranked item for IDM or most probable code at each head for CMM. Their entropy-matched smoothed counterparts distribute the remaining mass uniformly, matching the training-set mean entropy of the original targets. All four underperform full-distribution matching in R@10 and N@10 on every dataset. 
These comparisons show that retaining the full teacher distribution is more effective than using only its top choice or applying entropy-matched uniform smoothing. The gains therefore support the value of the teacher's nonuniform probability assignments beyond the highest-probability target.

\paragraph{Coverage of parallel code heads}
The 16-head variant supervises the first residual level in every subspace, the 32-head variant supervises the first two, and full CMM covers all three levels; hard-label training covers all 48 heads throughout. Across all three datasets, R@10 increases monotonically as supervision expands from 16 to 48 heads, confirming that fine-grained residual levels provide useful distinctions for candidate retrieval. Full coverage also achieves the highest N@10 on Beauty and Toys, while on the sparser Sports dataset, N@10 saturates at the first two residual levels (0.0347 vs. 0.0341 for full CMM). This indicates that coarse-to-mid residual codes already capture most top-ranking distinctions under sparser interactions, whereas full-hierarchy supervision remains consistently beneficial for broader recall.

\subsection{Sensitivity Analysis (RQ3)}
\label{sec:exp-sensitivity}
Having established the value of distributional targets, we examine teacher training data, target concentration, and distillation strength. Unless a hyperparameter is the factor being varied, sensitivity experiments retain the validation-selected student hyperparameters and calibration scales from Section~\ref{sec:exp-setup}.

\paragraph{Effect of teacher training data}
To examine how teacher training supervision affects knowledge transfer, we use SASRec teachers trained on nested 25\%, 50\%, and 100\% subsets of the training supervision rows. Appendix~\ref{app:teacher-data} specifies the sampling protocol.

\begin{table}[!htbp]
\centering
\caption{Effect of teacher training data (mean performance over three random seeds). Bold marks the highest performance across data fractions within each system and metric. Learning-rate-matched hard-label baselines ($r=0$) have R@10 and N@10 of 0.0940 and 0.0547 (Beauty, IDM), 0.0912 and 0.0537 (Beauty, CMM), 0.1000 and 0.0583 (Toys), and 0.0573 and 0.0322 (Sports).}
\label{tab:exp-teacher}
\small
\setlength{\tabcolsep}{2pt}
\begin{tabular*}{\linewidth}{@{\extracolsep{\fill}}llrrrrrr@{}}
\toprule
System & Data & \multicolumn{2}{c}{Beauty} & \multicolumn{2}{c}{Toys} & \multicolumn{2}{c}{Sports} \\
 & & R@10 & N@10 & R@10 & N@10 & R@10 & N@10 \\
\midrule
IDM & 25\% & $0.0934$ & $0.0479$ & $0.1065$ & $0.0549$ & $0.0577$ & $0.0284$ \\
 & 50\% & $0.1037$ & $0.0534$ & $0.1097$ & $0.0572$ & $0.0610$ & $0.0302$ \\
 & 100\% & $\mathbf{0.1094}$ & $\mathbf{0.0590}$ & $\mathbf{0.1176}$ & $\mathbf{0.0638}$ & $\mathbf{0.0644}$ & $\mathbf{0.0333}$ \\
\midrule
CMM & 25\% & $0.0962$ & $0.0511$ & $0.1076$ & $0.0589$ & $0.0585$ & $0.0297$ \\
 & 50\% & $0.1025$ & $0.0547$ & $0.1100$ & $0.0607$ & $0.0618$ & $0.0321$ \\
 & 100\% & $\mathbf{0.1069}$ & $\mathbf{0.0592}$ & $\mathbf{0.1144}$ & $\mathbf{0.0647}$ & $\mathbf{0.0639}$ & $\mathbf{0.0341}$ \\
\bottomrule
\end{tabular*}
\end{table}
Increasing teacher training supervision from 25\% to 100\% improves mean R@10 and N@10 for both IDM and CMM on every dataset (Table~\ref{tab:exp-teacher}). The consistent gains across the two matching spaces support the value of richer teacher supervision: additional teacher training data can benefit students even when their own training histories are unchanged.

\paragraph{Teacher temperature}
Unlike changing the teacher training data, temperature scaling modifies the concentration of the teacher's item probabilities while preserving its item ranking. We evaluate $\tau_T\in\{0.5,1,2,4\}$ using each dataset's full-data SASRec checkpoint.

\begin{table}[!htbp]
\centering
\caption{Teacher-temperature sensitivity (mean performance over three random seeds). Bold marks the best temperature per objective and metric.}
\label{tab:exp-temperature}
\small
\setlength{\tabcolsep}{2pt}
\begin{tabular*}{\linewidth}{@{\extracolsep{\fill}}llrrrrrr@{}}
\toprule
System & $\tau_T$ & \multicolumn{2}{c}{Beauty} & \multicolumn{2}{c}{Toys} & \multicolumn{2}{c}{Sports} \\
 & & R@10 & N@10 & R@10 & N@10 & R@10 & N@10 \\
\midrule
IDM & 0.5 & $0.1077$ & $0.0559$ & $0.1171$ & $0.0620$ & $0.0633$ & $0.0299$ \\
 & 1 & $\mathbf{0.1094}$ & $\mathbf{0.0590}$ & $\mathbf{0.1176}$ & $\mathbf{0.0638}$ & $\mathbf{0.0644}$ & $\mathbf{0.0333}$ \\
 & 2 & $0.0980$ & $0.0570$ & $0.1060$ & $0.0619$ & $0.0583$ & $0.0326$ \\
 & 4 & $0.0954$ & $0.0556$ & $0.1017$ & $0.0590$ & $0.0576$ & $0.0324$ \\
\midrule
CMM & 0.5 & $0.1027$ & $0.0551$ & $0.1130$ & $0.0618$ & $0.0602$ & $0.0294$ \\
 & 1 & $\mathbf{0.1069}$ & $\mathbf{0.0592}$ & $\mathbf{0.1144}$ & $\mathbf{0.0647}$ & $\mathbf{0.0639}$ & $\mathbf{0.0341}$ \\
 & 2 & $0.0912$ & $0.0532$ & $0.1013$ & $0.0587$ & $0.0573$ & $0.0321$ \\
 & 4 & $0.0881$ & $0.0517$ & $0.0989$ & $0.0574$ & $0.0548$ & $0.0305$ \\
\bottomrule
\end{tabular*}
\end{table}
The unscaled teacher distribution ($\tau_T=1$) yields the highest mean R@10 and N@10 for both objectives on all three datasets (Table~\ref{tab:exp-temperature}). Sharpening to $\tau_T=0.5$ concentrates mass on the highest-scoring items, whereas smoothing to $\tau_T\in\{2,4\}$ flattens relative probabilities. The results favor the unscaled distribution over either stronger concentration or greater uniformity. Because temperature preserves the teacher's ranking, this comparison highlights the role of probability allocation: teachers with identical item orderings can provide different supervision for parallel code prediction.

\paragraph{Sensitivity to distillation strength}
The preceding experiments change the teacher targets; we now vary their influence on student training. Using full-data teacher probabilities at $\tau_T=1$, we sweep nominal strength $r$ through $\lambda_m(r)=rC_m$. Appendix~\ref{app:strength} gives the evaluation grid.

\begin{figure}[!htbp]
  \centering
  \includegraphics[width=\linewidth]{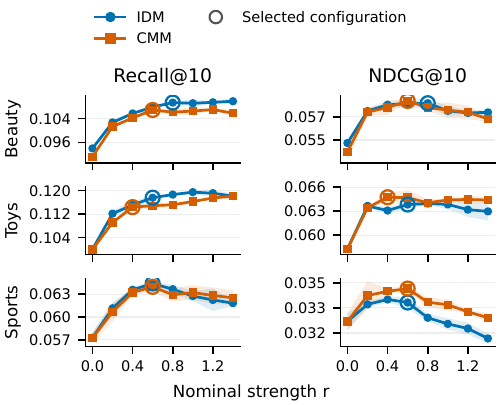}
  \caption{Distillation-strength sensitivity. Bands denote $\pm$ one sample standard deviation over three runs, and rings indicate the validation-selected configurations. The $r=0$ controls match the learning rate of each curve and coincide on Toys and Sports.}
  \label{fig:exp-strength}
  \Description{Rows correspond to Beauty, Toys, and Sports; columns show Recall at 10 and NDCG at 10 across eight strengths. Recall benefits persist over positive strengths, while NDCG can decline at large strengths, particularly on Sports.}
\end{figure}

Every positive strength improves mean R@10 over its matched $r=0$ control (Figure~\ref{fig:exp-strength}), showing that retrieval benefits extend across the weight range. NDCG is more sensitive: moderate supervision generally performs well, while excessive strength can reduce top-ranked accuracy. The objectives also respond differently; on Toys, CMM's N@10 peaks at a lower strength than IDM's, even though both retain Recall gains at larger weights. This behavior connects the weight of distribution matching to the retrieval--ordering distinction observed in RQ1 and motivates validation-based selection of the distillation strength.

\subsection{Joint Training and Prediction Fusion (RQ4)}
\label{sec:exp-combination}
\paragraph{Combination schemes}
Finally, we compare two ways to combine the matching spaces: \emph{in-training optimization mixing} (Joint), which applies both distillation losses to one student alongside hard-label supervision, and \emph{post-hoc score-level ensembling} (Fusion), which averages the catalogue ranking scores of separately trained IDM and CMM students. For Fusion, we pair students by initialization seed and use equal weights of 0.5/0.5, combining two predictors without retraining. Appendix~\ref{app:combination} specifies the joint objective, validation selection, and fusion rule.

\paragraph{Comparison and findings}
Joint training mixes a shared teacher-marginal signal with distinct catalogue-level and head-wise normalization terms, as characterized in Section~\ref{sec:matching_relationship}. The results favor retaining separately optimized predictors and combining their catalogue scores: prediction fusion attains the highest mean N@10 on all three datasets and the highest mean R@10 on Toys and Sports (Table~\ref{tab:exp-combine}). These gains provide empirical evidence that IDM and CMM learn complementary predictions whose combination can improve upon either standalone model.

\begin{table}[!htbp]
\centering
\caption{Standalone IDM/CMM, joint training, and same-seed score fusion (mean performance over three random seeds). Bold denotes the highest unrounded mean for each dataset and metric.}
\label{tab:exp-combine}
\small
\setlength{\tabcolsep}{2pt}
\begin{tabular*}{\linewidth}{@{\extracolsep{\fill}}lrrrr@{}}
\toprule
System & R@5 & R@10 & N@5 & N@10 \\
\midrule
\multicolumn{5}{l}{\emph{Beauty}} \\
IDM & $0.0719$ & $\mathbf{0.1094}$ & $0.0468$ & $0.0590$ \\
CMM & $0.0714$ & $0.1069$ & $0.0478$ & $0.0592$ \\
Joint & $0.0711$ & $0.1069$ & $0.0470$ & $0.0586$ \\
Fusion & $\mathbf{0.0728}$ & $0.1093$ & $\mathbf{0.0479}$ & $\mathbf{0.0597}$ \\
\midrule
\multicolumn{5}{l}{\emph{Toys}} \\
IDM & $0.0804$ & $0.1176$ & $0.0518$ & $0.0638$ \\
CMM & $0.0784$ & $0.1144$ & $0.0532$ & $0.0647$ \\
Joint & $0.0795$ & $0.1178$ & $0.0515$ & $0.0638$ \\
Fusion & $\mathbf{0.0815}$ & $\mathbf{0.1184}$ & $\mathbf{0.0536}$ & $\mathbf{0.0655}$ \\
\midrule
\multicolumn{5}{l}{\emph{Sports}} \\
IDM & $0.0423$ & $0.0644$ & $0.0262$ & $0.0333$ \\
CMM & $0.0422$ & $0.0639$ & $0.0272$ & $0.0341$ \\
Joint & $0.0420$ & $0.0633$ & $0.0268$ & $0.0337$ \\
Fusion & $\mathbf{0.0433}$ & $\mathbf{0.0659}$ & $\mathbf{0.0272}$ & $\mathbf{0.0344}$ \\
\bottomrule
\end{tabular*}
\end{table}

\FloatBarrier
\section{Conclusion}
\label{sec:conclusion}

Cross-space distillation enriches hard-label supervision for parallel semantic-ID recommendation. IDM and CMM share teacher code marginals through fixed SID assignments but normalize student predictions differently: catalogue-level competition and direct head-wise alignment. These complementary signals improve recommendation while preserving parallel inference.

Experiments on three benchmark datasets show that both objectives outperform the compared state-of-the-art baselines. IDM favors Recall and candidate coverage, whereas CMM favors NDCG and top-ranked ordering across all three datasets, providing a practical choice between the two matching spaces. Score fusion further improves NDCG across all three datasets by combining their complementary predictions. Future work will explore joint distillation with learnable item tokenizers, step-wise distribution alignment for masked or diffusion-based recommenders, and adaptive matching strategies that balance retrieval coverage with ranking quality.

\clearpage
\bibliographystyle{ACM-Reference-Format}
\bibliography{references}

\clearpage
\appendix
\section{Experimental and Implementation Details}
\label{app:implementation}
\label{app:reproducibility}

\subsection{Data Processing and Supervision}
The input sequences are the 5-core Amazon Reviews 2014 datasets. We verify the 5-core condition without applying further filtering or deduplication. For each user, interactions are ordered chronologically; events sharing a timestamp retain their original order. The final and penultimate interactions are used as the Test and Validation targets, respectively. Earlier events provide training supervision.

The training procedure constructs 25,745, 22,181, and 37,197 sliding windows for Beauty, Toys, and Sports, respectively. Each user's initial window contains at most 51 events and supervises every valid next-item position; subsequent windows supervise only their last target, thereby avoiding repeated supervision of the same position. Histories are limited to 50 items, and Test histories include the observed Validation interaction. Distillation matches each supervised position to its exact historical prefix.

Each of the 48 student heads predicts a distribution over the 256 code classes directly, without requiring auxiliary special tokens (such as PAD or EOS) in the output vocabulary. Evaluation ranks all valid catalogue items without removing items previously observed in the user's history. Fixed complete SIDs are not necessarily unique: Beauty, Toys, and Sports contain 3, 3, and 25 collision groups, comprising 6, 12, and 67 items, respectively. Colliding items receive identical scores under a given history, but each item remains a distinct entry in catalogue normalization.

\subsection{Student and Teacher Configurations}
\paragraph{Parallel student.}
The student uses the AsymRec architecture~\cite{huang2026asymrec}, with fixed 3,072-dimensional \texttt{text-embedding-3-large} item representations. The Kronecker residual bridge maps $48\times64$ input dimensions to $16\times28$ dimensions, resulting in a hidden dimension of 448. The sequence encoder contains two Transformer layers, four attention heads, and a feed-forward dimension of 1,024. The maximum history length is 50. The fixed SID assignment uses 16 quantization subspaces with three residual-quantization levels each, yielding 48 categorical output heads of 256 classes. The effective head logits incorporate the student temperature of 0.03.

Embedding and attention dropout rates are 0.5 for Beauty and Sports and 0.3 for Toys, while residual dropout is disabled. Layer-normalization epsilon is $10^{-5}$ in the bridge and $10^{-12}$ in the backbone; the parameter-initialization standard deviation is 0.02. Weight decay is $10^{-5}$, $10^{-3}$, and 0 for Beauty, Toys, and Sports, respectively. Student training uses AdamW with batch size 256 for up to 150 epochs, 10,000 warm-up updates followed by cosine learning-rate decay, and gradient clipping at norm 1. Checkpoints are selected by Validation N@10 with early-stopping patience 20. Computation uses FP32; the evaluation batch size is 32, and training is performed on RTX 5880 Ada GPUs.

\paragraph{Sequential teacher.}
The SASRec~\cite{kang2018sasrec} teacher has hidden dimension 256, two causal self-attention layers with two heads, feed-forward dimension 1,024, and maximum history length 50. On Beauty and Toys, dropout is 0.3, weight decay is $10^{-5}$, and the Adam learning rates are approximately $1.395\times10^{-4}$ and $1.197\times10^{-4}$, respectively. Sports uses dropout 0.2, learning rate $10^{-3}$, and zero weight decay. Teacher training uses batch size 512, a constant learning rate, at most 200 epochs, and early-stopping patience 20, without gradient clipping. The teacher is trained with full-catalogue cross-entropy, and its logits over the N valid catalogue items are normalized via Eq. (4) at $\tau_T$ = 1 to produce the distillation targets.

\subsection{Gradient Calibration and Hyperparameter Selection}
\label{app:calibration}
To account for differences between hard-label and auxiliary gradient magnitudes, we calibrate the initial auxiliary scale separately for each dataset and matching objective using a separately initialized student model. After ten hard-label training batches, we compute the full-trainable-parameter $L_2$ norms of the hard-label and auxiliary gradients over the following twenty batches. The scale is the median of the per-batch norm ratios:
\begin{equation}
 C_m=\operatorname{median}_{b=11,\ldots,30}
 \frac{\|\nabla_\theta\mathcal L_{\mathrm{hard}}^{(b)}\|_2}
 {\|\nabla_\theta\mathcal L_m^{(b)}\|_2},\qquad
 \lambda_m=r_mC_m.
\label{eq:gradient_calibration}
\end{equation}
Here, $r_m$ controls the nominal distillation strength. Gradient norms cover all trainable student parameters. The resulting $\lambda_m$ is fixed across all three runs for each selected dataset--objective configuration.

The linear ramp in Eq.~\eqref{eq:training_objective} is $w(t)=\min\{t/R,1\}$, where $R$ is the number of optimizer updates over which distillation increases from zero to its full weight. It lasts approximately ten epochs; Table~\ref{tab:app-recipes} lists the exact update counts.

Learning-rate candidates are 0.005, 0.01, and 0.02, with 0.0025 additionally considered on Beauty. Nominal distillation strengths are selected from $\{0.2,0.4,0.6,0.8\}$. Learning rates and strengths are chosen through a joint assessment of Validation ranking performance; within each run, the checkpoint with the highest Validation N@10 is retained. The selected configurations are fixed for evaluations over three random seeds. A superscript $\star$ denotes the validation-selected setting for objective $m$.

\begin{table}[htbp]
\centering
\caption{Selected single-model hyperparameters. All three runs use the same dataset--objective coefficient.}
\label{tab:app-recipes}
\small
\setlength{\tabcolsep}{2pt}
\begin{tabular*}{\linewidth}{@{\extracolsep{\fill}}llrrrr@{}}
\toprule
Dataset & Method & LR & $r_m^\star$ & $\lambda_m^\star$ & Ramp steps \\
\midrule
Beauty & AsymRec & 0.005 & --- & --- & --- \\
Beauty & IDM & 0.005 & 0.8 & 1.716 & 1048 \\
Beauty & CMM & 0.02 & 0.6 & 0.802 & 1048 \\
Toys & AsymRec & 0.02 & --- & --- & --- \\
Toys & IDM & 0.01 & 0.6 & 1.530 & 903 \\
Toys & CMM & 0.01 & 0.4 & 0.493 & 903 \\
Sports & AsymRec & 0.01 & --- & --- & --- \\
Sports & IDM & 0.01 & 0.6 & 1.962 & 1515 \\
Sports & CMM & 0.01 & 0.6 & 0.980 & 1515 \\
\bottomrule
\end{tabular*}
\end{table}

\begin{table*}[!t]
\centering
\caption{Four-metric ablation results (mean performance over three random seeds). Variant names and objective groups follow Table~\ref{tab:exp-ablation}.}
\label{tab:app-ablation}
\small
\setlength{\tabcolsep}{2pt}
\begin{tabular*}{\linewidth}{@{\extracolsep{\fill}}lrrrrrrrrrrrr@{}}
\toprule
Variant & \multicolumn{4}{c}{Beauty} & \multicolumn{4}{c}{Toys} & \multicolumn{4}{c}{Sports} \\
\cmidrule(lr){2-5}\cmidrule(lr){6-9}\cmidrule(lr){10-13}
 & R@5 & R@10 & N@5 & N@10 & R@5 & R@10 & N@5 & N@10 & R@5 & R@10 & N@5 & N@10 \\
\midrule
IDM (full) & $0.0719$ & $0.1094$ & $0.0468$ & $0.0590$ & $0.0804$ & $0.1176$ & $0.0518$ & $0.0638$ & $0.0423$ & $0.0644$ & $0.0262$ & $0.0333$ \\
Top-1 item & $0.0687$ & $0.1041$ & $0.0434$ & $0.0548$ & $0.0773$ & $0.1141$ & $0.0489$ & $0.0608$ & $0.0387$ & $0.0606$ & $0.0223$ & $0.0294$ \\
Uniform-smoothed item & $0.0688$ & $0.1038$ & $0.0435$ & $0.0548$ & $0.0782$ & $0.1144$ & $0.0495$ & $0.0612$ & $0.0389$ & $0.0602$ & $0.0224$ & $0.0293$ \\
\midrule
CMM (full) & $0.0714$ & $0.1069$ & $0.0478$ & $0.0592$ & $0.0784$ & $0.1144$ & $0.0532$ & $0.0647$ & $0.0422$ & $0.0639$ & $0.0272$ & $0.0341$ \\
Hard code & $0.0677$ & $0.1010$ & $0.0438$ & $0.0545$ & $0.0767$ & $0.1119$ & $0.0503$ & $0.0616$ & $0.0376$ & $0.0595$ & $0.0221$ & $0.0292$ \\
Uniform-smoothed code & $0.0681$ & $0.1006$ & $0.0443$ & $0.0547$ & $0.0763$ & $0.1103$ & $0.0507$ & $0.0617$ & $0.0388$ & $0.0599$ & $0.0237$ & $0.0305$ \\
16 heads (level 1) & $0.0693$ & $0.1011$ & $0.0481$ & $0.0584$ & $0.0761$ & $0.1095$ & $0.0531$ & $0.0639$ & $0.0410$ & $0.0613$ & $0.0281$ & $0.0346$ \\
32 heads (levels 1--2) & $0.0711$ & $0.1046$ & $0.0483$ & $0.0591$ & $0.0774$ & $0.1112$ & $0.0534$ & $0.0643$ & $0.0416$ & $0.0635$ & $0.0277$ & $0.0347$ \\
\bottomrule
\end{tabular*}
\end{table*}

\begin{table*}[!t]
\centering
\caption{Gradient-calibrated coefficients and entropy-matched smoothing parameters used by the ablation variants. Values are fixed across runs.}
\label{tab:app-ablation-weights}
\small
\setlength{\tabcolsep}{2pt}
\begin{tabular*}{\linewidth}{@{\extracolsep{\fill}}lrrrrrr@{}}
\toprule
Variant & \multicolumn{2}{c}{Beauty} & \multicolumn{2}{c}{Toys} & \multicolumn{2}{c}{Sports} \\
\cmidrule(lr){2-3}\cmidrule(lr){4-5}\cmidrule(lr){6-7}
 & $\lambda$ & $\epsilon$ & $\lambda$ & $\epsilon$ & $\lambda$ & $\epsilon$ \\
\midrule
Top-1 item & 0.867649 & --- & 0.779020 & --- & 0.611078 & --- \\
Uniform-smoothed item & 1.462578 & 0.406484 & 1.240714 & 0.371723 & 1.690206 & 0.638363 \\
\midrule
Hard code & 0.542772 & --- & 0.379317 & --- & 0.528670 & --- \\
Uniform-smoothed code & 0.853681 & 0.487760 & 0.518572 & 0.465291 & 1.146013 & 0.752142 \\
16 heads (level 1) & 0.444682 & --- & 0.292024 & --- & 0.548509 & --- \\
32 heads (levels 1--2) & 0.654976 & --- & 0.408500 & --- & 0.793183 & --- \\
\bottomrule
\end{tabular*}
\end{table*}

\subsection{RPG Implementation and Evaluation}
\label{app:rpg}
The local RPG comparison uses the authors' \href{https://github.com/facebookresearch/RPG_KDD2025/blob/7dcf95ce9e08601810efdd1c850ed6dada5f97c2/README.md}{released implementation} (revision \texttt{7dcf95ce9e08}) and its graph-guided decoding~\cite{hou2025rpg}. Its SID lengths are 32, 16, and 16 for Beauty, Toys, and Sports, and the corresponding learning rates are 0.01, 0.003, and 0.003. The logit temperature is 0.03. We use released decoding parameters $(b,k,q)$ of $(20,200,3)$, $(200,20,3)$, and $(100,30,5)$, respectively. Table~\ref{tab:exp-main} reports the resulting RPG performance.

\section{Additional Ablation Results}
\label{app:ablations}

\paragraph{Soft-target variants.}
For IDM, \emph{Top-1 item} places a one-hot target $e$ at the teacher's highest-probability valid item; \emph{Uniform-smoothed item} uses $(1-\epsilon)e+\epsilon\mathbf 1_N/N$. For CMM, \emph{Hard code} selects the most probable code at each head, and \emph{Uniform-smoothed code} uses $(1-\epsilon)e+\epsilon\mathbf 1_K/K$. Here, $\mathbf 1_N$ and $\mathbf 1_K$ are all-ones vectors over catalogue items and code values, respectively. Each smoothing parameter matches the training-set mean entropy of the corresponding original teacher targets, averaged over heads for code targets. Independently selected hard codes need not form the SID of a valid catalogue item.

\paragraph{Partial-head variants.}
The 48 code positions are organized as 16 quantization subspaces with three residual-quantization levels per subspace. The \emph{16-head} variant distills the first residual level in every subspace, using zero-based head indices $\{3m:m=0,\ldots,15\}$; the \emph{32-head} variant distills the first two levels, using $\{3m,3m+1:m=0,\ldots,15\}$. The CMM loss averages over the 16 or 32 active heads, respectively, while the original hard-label objective supervises all 48 heads.

All ablations retain the parent objective's learning rate and nominal distillation strength, with a separate gradient calibration for the modified auxiliary target or head subset. The resulting coefficients and smoothing parameters are listed in Table~\ref{tab:app-ablation-weights}.

Table~\ref{tab:app-ablation} complements the main ablation table with results at cutoff 5, including the full IDM/CMM references for all four metrics.

\section{Additional Sensitivity Results}
\label{app:teacher-strength}

\subsection{Teacher Training Data}
\label{app:teacher-data}
Teacher training data are sampled from history--target supervision rows using a fixed Seed-42 permutation within each domain. The 25\% and 50\% subsets are nested and preserve each row's full preceding history. Students always use the complete training data and retain the distillation coefficient selected with the full-data teacher. The Sports subsets contain 47,385 and 94,771 supervised target positions at 25\% and 50\%, respectively. All teachers use a common SASRec architecture, item mapping, and catalogue evaluation protocol.

\subsection{Distillation Strength}
\label{app:strength}
We evaluate $r\in\{0,0.2,\ldots,1.4\}$ with the full-data teacher fixed at $\tau_T=1$. Each curve retains the calibration scale $C_m$ of its validation-selected configuration, giving
\begin{equation}
  \lambda_m(r)=rC_m=\lambda_m^\star\frac{r}{r_m^\star}.
  \label{eq:exp-strength}
\end{equation}
Figure~\ref{fig:exp-strength} reports Recall@10 and NDCG@10 across this strength grid.

\section{Joint Training and Score Fusion}
\label{app:combination}

\subsection{Joint Training Configuration}
The joint method uses one student forward pass and applies the hard-label loss once. Its auxiliary objective is
\begin{equation}
\begin{aligned}
\mathcal L_{\mathrm{aux,joint}}(x;t)
&=r_JC_Jw(t)\bigl[\alpha C_M\mathcal L_{\mathrm{CMM}}(x)\\
&\qquad +(1-\alpha)C_I\mathcal L_{\mathrm{IDM}}(x)\bigr],
\end{aligned}
\label{eq:joint_auxiliary}
\end{equation}
where $C_M$ and $C_I$ are the calibration scales of the independent matching objectives. The joint scale $C_J$ is calibrated using the gradient of their combined auxiliary loss, which includes the interaction between the two gradients. Candidate mixture weights are $\alpha\in\{0.25,0.5,0.75\}$ and strengths are $r_J\in\{0.4,0.6,0.8\}$. Learning-rate candidates are 0.005 and 0.02 on Beauty, and 0.005 and 0.01 on Toys and Sports. Hyperparameters are selected through a joint assessment of Validation ranking performance. The selected learning rate is 0.005 on all three datasets; the coefficients in Table~\ref{tab:app-joint-config} remain fixed across runs.

\begin{table}[t]
\centering
\caption{Selected joint-training configurations. The final columns give the effective post-ramp weights for CMM and IDM, respectively.}
\label{tab:app-joint-config}
\small
\setlength{\tabcolsep}{2pt}
\begin{tabular*}{\linewidth}{@{\extracolsep{\fill}}lrrrr@{}}
\toprule
Dataset & $\alpha$ & $r_J$ & $\lambda_{\mathrm{CMM}}$ & $\lambda_{\mathrm{IDM}}$ \\
\midrule
Beauty & 0.75 & 0.8 & 0.859846 & 0.460198 \\
Toys & 0.25 & 0.6 & 0.205164 & 1.274258 \\
Sports & 0.50 & 0.4 & 0.373973 & 0.748308 \\
\bottomrule
\end{tabular*}
\end{table}

\subsection{Same-seed Score Fusion}
The main-paper fusion combines separately trained IDM and CMM students with the same seed. It averages the catalogue ranking scores computed from the mean log-probability of each item's 48 assigned codes:
\begin{equation}
 s_{\mathrm{fusion}}(i\mid x)
 =\tfrac12 s_{\mathrm{IDM}}(i\mid x)
 +\tfrac12 s_{\mathrm{CMM}}(i\mid x).
\end{equation}
Fusion directly uses raw scores at inference without calibration or tuning.

\FloatBarrier

\section{Mathematical Details of Item--Code Matching}
\label{app:mathematical_details}

We fix a history $x$ and omit its conditioning below. Teacher probabilities and SID assignments are constants with respect to the student parameters. All logarithms are natural, with $0\log0=0$. The identities concern the unweighted losses for one supervised history; minibatch averaging, the fixed coefficient, and the ramp are applied subsequently.

\subsection{Equivalent Catalogue Scores}
\label{app:equivalent_scores}

The catalogue-level and head-averaged log-partition terms used below are
\begin{equation}
\begin{aligned}
A(x)&=\log\sum_{i\in\mathcal I}\exp(s_i(x)),\\
B(x)&=\frac1H\sum_{h=1}^{H}
\log\sum_{c\in\mathcal C}\exp(\ell_{h,c}(x)),\\
\Delta_{\mathrm{norm}}(x)&=A(x)-B(x).
\end{aligned}
\label{eq:appendix_partition_terms}
\end{equation}
Define $Z_h=\sum_c\exp(\ell_{h,c})$. The evaluation score satisfies
\begin{equation}
\begin{aligned}
\widetilde s_i
&=\frac1H\sum_h\log q_h(z_{i,h})\\
&=\frac1H\sum_h\ell_{h,z_{i,h}}
  -\frac1H\sum_h\log Z_h
=s_i-B.
\end{aligned}
\end{equation}
Since $B$ is independent of $i$, both scoring rules yield the same ranking and catalogue softmax.

\subsection{Proof of Proposition 3.1: Forward-KL Decomposition}
Because every row of $M_h$ contains exactly one 1, $\mathbf b_h=M_h^\top\mathbf p_T$ is a probability vector. Let $P=P_T^Z$ denote the teacher pushforward on complete code tuples, whose marginals are $b_h$. For $Q_\theta^Z(\mathbf z)=\prod_hq_h(z_h)$,
\begin{equation}
\begin{aligned}
D_{\mathrm{KL}}(P\Vert Q_\theta^Z)
&=-\mathcal H(P)-\sum_h\sum_c b_h(c)\log q_h(c),\\
\sum_h D_{\mathrm{KL}}(b_h\Vert q_h)
&=-\sum_h\mathcal H(b_h)
  -\sum_h\sum_c b_h(c)\log q_h(c).
\end{aligned}
\end{equation}
Their difference is
\begin{equation}
\sum_h\mathcal H(b_h)-\mathcal H(P)
=D_{\mathrm{KL}}\!\left(P\,\middle\Vert\,\prod_hb_h\right)
=\operatorname{TC}(P).
\end{equation}
This proves Eq.~\eqref{eq:factorized_forward_kl}, which holds generally regardless of SID collisions or inter-position dependencies in the teacher distribution.

\subsection{Proof of Proposition 3.2: Loss and Gradient Identities}
The teacher-weighted additive score depends only on its code marginals:
\begin{equation}
\sum_{i\in\mathcal I}p_T(i\mid x)s_i(x)
=\frac1H\sum_{h=1}^{H}\sum_{c\in\mathcal C}
 b_h(c\mid x)\ell_{h,c}(x).
\label{eq:shared_teacher_score}
\end{equation}
Writing $A$ and $B$ as in Eq.~\eqref{eq:appendix_partition_terms}, the losses expand to
\begin{equation}
\begin{aligned}
\mathcal L_{\mathrm{IDM}}
&=-\frac1H\sum_{h,c}b_h(c)\ell_{h,c}+A,\\
\mathcal L_{\mathrm{CMM}}
&=-\frac1H\sum_{h,c}b_h(c)\ell_{h,c}
  +B-\frac1H\sum_h\mathcal H(b_h).
\end{aligned}
\end{equation}
Subtracting gives the loss identity in Eq.~\eqref{eq:normalization_identity}. The teacher entropies have zero student gradient, leaving $\nabla_\theta(A-B)$ as the correction between the two parameter gradients.

For effective head logits, $\partial s_i/\partial\ell_{h,c}=(M_h)_{i,c}/H$. Hence
\begin{equation}
\begin{aligned}
\frac{\partial A}{\partial\ell_{h,c}}
&=\frac1H\sum_iq_{\mathrm{item}}(i)(M_h)_{i,c}
=\frac{\bar q_h(c)}H,\\
\frac{\partial B}{\partial\ell_{h,c}}
&=\frac{q_h(c)}H.
\end{aligned}
\end{equation}
Combining these derivatives with the common linear term proves Eq.~\eqref{eq:matching_logit_gradients}.

\clearpage

\end{document}